\documentclass[reprint]{revtex4-2}

\usepackage{graphicx}
\usepackage{dcolumn}
\usepackage{bm}

\begin{document}

\title{Dual-wavelength control of charge accumulation in rubrene microcrystals with anisotropic conductivity}

\author{Moha Naeimi}
\affiliation{Institute of physics, University of Rostock, Germany}
\affiliation{Department of light, life and matter, University of Rostock, Germany}

\author{Ingo Barke}
\affiliation{Institute of physics, University of Rostock, Germany}
\affiliation{Department of light, life and matter, University of Rostock, Germany}

\author{Sylvia Speller}
\email{corresponding author email: sylvia.speller@uni-rostock.de}
\affiliation{Institute of physics, University of Rostock, Germany}
\affiliation{Department of light, life and matter, University of Rostock, Germany}

\date{\today}

\begin{abstract}
Previously, a novel type of rubrene microcrystals was reported, forming two distinct sectors—diamond- and triangular-shaped—that exhibit pronounced contrasts in photoluminescence (PL) spectra and exciton dynamics. In the present work, their internal electronic structure is investigated using time-of-flight photoemission electron spectroscopy (TOF-PES), revealing that the two sector's different charging characteristics arising from anisotropic conductivities. Upon photoemission via a one-photon photoemission (1PPE) process excited by 6.2 eV (200 nm) photons, the diamond-shaped sectors accumulate significant charge, whereas the triangular sectors remain essentially uncharged. The charge accumulation in the diamond sectors can be neutralized by additional sub-threshold illumination, which generates charge carriers through internal photoeffect. The dynamics and energetics of the observed band shifting is described quantitatively by a model combining surface capacitance and drift-diffusion. These crystalline systems enable the creation of built-in charge landscapes that can be manipulated both spatially and temporally.

\end{abstract}
\maketitle

\section{Introduction}
Organic crystals and aggregates have gained significant attention in organic electronics and materials science due to their unique electronic, optical, and structural properties. With a tetracene backbone and 4 phenyl groups, rubrene is subject to the huge interest of research with numerous applications in electronic devices ranging from field effect transistors \cite{Kim2007, Jo2015} to light emitting diodes \cite{Wang2023} offering one of the highest charge mobilities \cite{Zhang2010} among organic molecules and a very efficient singlet fission in the crystalline form. Through singlet fission \cite{Ma2012, Liu2023, Bardeen2012} at $\sim$ 530 nm, long lived triplet excitons are generated, which can diffuse extensively within the bulk as well as along the surface playing a crucial role in energy transport and interfacial processes \cite{Euvrard2022}. Furthermore, its strong absorption in the visible range and favorable energy level alignment with common acceptor materials have driven extensive research into rubrene-based solar cells \cite{AkinKara2022, Yang2018}. Despite these advantages, the photovoltaic performance of rubrene remains highly dependent on crystal quality, morphology, and interface engineering, highlighting the need for controlled growth strategies to fully exploit its potential in next-generation organic photovoltaics.

For efficient charge and exciton transfer, the morphology of the molecular and lattice structure plays a critical role. Previous studies have shown that this process is strongly suppressed in the amorphous phase of rubrene \cite{Finton2019, Takahashi2019}. The remaining challenge, therefore, lies in preparing large, single-domain, and flat crystals in the orthorhombic phase \cite{Clapham2021}, where the singlet fission is most efficient \cite{Wakikawa2025, Ma2013}. Various strategies may be employed to address this challenge, including thermal treatment \cite{Fielitz2016} and crystal growth under extreme conditions such as physical vapour transport (PVT) \cite{Zeng2007} or high partial pressure \cite{Ye2018}.

We recently reported a novel type of rubrene orthorhombic crystals \cite{Naeimi2025} growing under high partial pressure and elevated temperature. These crystals tend to have either the \textbf{c}- or \textbf{b}-axis of the orthorhombic unit cell as the out-of-plane axis. The out-of-plane axes significantly impact the PL spectra and exciton dynamics arising from the orientation of the rubrene transition dipole moment, which is aligned along the \textbf{c}-axis of the unit cell. In both growth directions, a particular growth habit was observed exhibiting 4 sectors of two distinct types within one single crystal, determining PL spectra and exciton dynamics. The source of this zone-sectoring is the rotation of the orthorhombic unit cell around axes during the growth impacting the absorption rates and exciton diffusion direction along the crystal in- and out-of- plane.

Understanding the fundamental electronic structure of rubrene single crystals is essential for investigating their charge transport and photo-voltaic properties. Angle-resolved photoelectron spectroscopy (ARPES) experiments showed that the highest occupied molecular orbital (HOMO) valence band in rubrene exhibits a dispersion of 0.4 eV along the strongly $\pi-\pi$-stacked crystallographic directions (\textbf{a}-axis in orthorhombic phase), supporting a band-like transport in rubrene \cite{Machida2010}. Subsequent wavelength dependent photoemission studies further resolved the two-dimensional valence band structure, revealing anisotropy in intermolecular transfer confirming the quasi-two-dimensional nature of the HOMO band \cite{Nakayama2008}. More recent angle-resolved measurements have shown that near-surface electronic doping via molecular electron donors and acceptors can shift the Fermi level within the band gap of rubrene without significantly perturbing the intrinsic valence band dispersion, indicating that charge injection and accumulation processes can be probed\cite{Li2019}. 

We present a photoemission study on the zone-sectored rubrene microcrystal, showing that the diamond-shaped sectors accumulate significant charge under a one-photon photoemission (1PPE) process when excited with photons above the work function, while the triangular sectors remain largely unaffected. Analysis of the charge dynamics indicates that the sectors exhibit strongly anisotropic conductivities and behave like micro-scale capacitors, characterized by a distinct time constant. The photoconductivity and subsequent charge could be controlled with another excitation with lower photon energies by generating electron-hole pairs resulting from internal photo-effect.

\section{Experimental methods}

\begin{figure}
    \centering
    \includegraphics[width=1\linewidth]{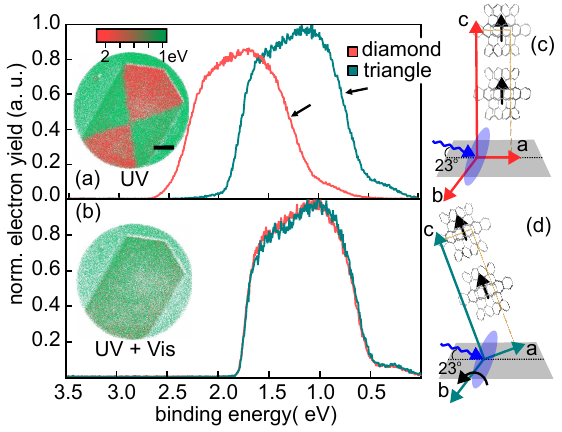}
    \caption{(a) 1PPE electron spectra of a rubrene crystal resolved for different sectors 6.2 eV illumination. (b) 1PPE electron spectra of the same crystal upon two colour illumination (6.2 eV and 3.1 eV). The insets are PEEM maps which colour-code the spatially-resolved centre of spectral weight. The charge patterns resemble the diamond- and triangular-shaped sectors. (scale bar: 10 $\mu$m) (c and d) Schematic of the orthorhombic \textbf{ac}-plane in diamond- and triangular-shaped sectors, respectively. The blue arrows and circles, represent the excitation light and polarization plane, applied with the angle of $23^\circ$ grazing. Note that the spectra for the diamond sector are shifted towards higher "apparent" binding energies.}
    \label{fig:fs_cw_img}
\end{figure}

The preparation method of the crystals is described in our previous work \cite{Naeimi2025}. For surface potential investigations we used an atomic force microscope (Park Systems NX20) in sideband Kelvin probe mode with conductive tips coated by a chromium platinum (Cr-Pt) layer exhibiting a cantilever spring constant of 3 N/m and a free eigen frequency of 75 kHz.

The time-of-flight photoemission electron spectroscopy was conducted in a PEEM (Focus IS-PEEM) by placing the sample to a vacuum chamber with a base pressure of $10^{-10}$ mbar. The light angle of incidence was 23 degrees grazing, ensuring a good alignment of the light polarisation and transition dipole moment of \textbf{c}-oriented rubrene crystals, which is perpendicular to the substrate.

We used two different light sources for photoemission and surface potential investigations: (1) The 4th harmonics of a tunable Ti:Sa femtosecond (fs) laser (Mira 900F) yielding photons with energies of 6.2 eV (200 nm) with the repetition rate of 1 Mhz and pulse duration of 200 ps. (2) A continues wave (cw) laser with photon energies of 3.09 eV (405 nm).

Visualization and analysis was done by Gwyddion \cite{Necas2012}and Igor Pro (Wavemetrics). Visualization and analysis was done by Gwyddion \cite{Necas2012}and Igor Pro (Wavemetrics). We calculated an apparent binding energy from the kinetic energy of the detected electrons assuming a 1-photon photoemission process. In case of charging, this “apparent” binding energy will not reflect the mere binding energy but additionally accounts for the reduction in the photo electron’s kinetic energy by the positive charging of the surface.

\section{Results and discussion}

We investigated a \textbf{c}-oriented rubrene microcrystal (see \cite{Naeimi2025}), in which the crystallographic \textbf{c}-axis i.e. the transition dipole moment of rubrene molecules in the orthorhombic phase is perpendicular to the surface. A key advantage of this crystal orientation is that, under grazing incidence, the polarization plane of the excitation light aligns well with the transition dipole moment, resulting in enhanced absorption if excitons are involved. The distinction between the diamond- and triangular-shaped sectors of this crystal arises from the rotation of the unit cell around the \textbf{b}-axis (see Figure \ref{fig:fs_cw_img}c and \ref{fig:fs_cw_img}d). In the triangular sectors, this rotation is more pronounced than in the diamond sectors, as shown in our previous work \cite{Naeimi2025}. Such rotation impacts the photoluminescence spectra and exciton dynamics. 

\begin{figure}
    \centering
    \includegraphics[width=1\linewidth]{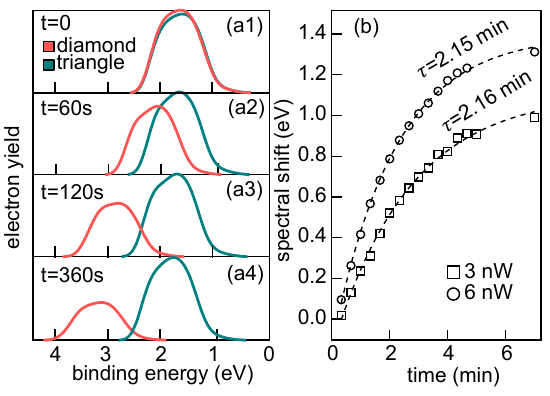}
    \caption{It takes some time for the crystal to fully charge. (a) Electron spectra of the crystal upon illumination (6.16 eV) for different times after illumination is started. (b) Shift of spectra (charge) versus time after illumination has started for two different illumination powers.}
    \label{fig:fs_spec_shift}
\end{figure}

Figure \ref{fig:fs_cw_img}a shows the electron spectra of a \textbf{c}-oriented crystal, resolved for different sectors. The excitation source is a 6.2 eV (200 nm) femtosecond pulsed laser (here we refer to it as "UV" light) operating at a repetition rate of 1 MHz with an average power of 2 nW. The electrons are emitted through a 1PPE process. The acquisition time of the spectra is 20 seconds, while the sample was already exposed to the illumination for more than 5 minutes. As shown, the spectra from the diamond sectors are shifted by approximately 0.75 eV toward higher binding energies, while the overall spectral shape remains unchanged. The different slopes at the high kinetic energies (marked with arrows in Figure \ref{fig:fs_cw_img}a) are mainly caused by a residual time-dependent shift of the spectrum within the relatively long data acquisition time.

Figure \ref{fig:fs_cw_img}b shows the electron spectra when an additional illumination source is simultaneously applied. The second illumination is a 3.06 eV (405 nm) continuous wave laser (here we refer to it as "vis" light) with the same power as the UV source. While no additional photoemission is induced by the Vis light, the spectra from different sectors are similar. We attribute the weak peak at lower binding energies to possible trap states caused by crystal defects or to the triplet state (1.14 eV) generated via singlet fission. The width of the band measures about 1 eV. This is wider than ARPES studies \cite{Vollmer2012}, however, we tentatively attribute the band to the HOMO band (highest occupied molecular orbital) having in mind our extraction lens collects electrons with all k-vectors.

As the entire spectra of the diamond sectors shift toward higher binding energies without altering the spectral shape, this behaviour can be attributed to charging which is more pronounced in diamond-shaped sectors compared to the triangular-shaped ones. In the photoemission process, charging is the result of hole accumulation within the semiconductor as photoelectrons are liberated. This accumulation of positively charged holes cannot be compensated by electron injection from the substrate and further electron emission, thereby resulting in a shift toward "apparently" higher binding energies \cite{Gilbert2000, Wilson2020}. 

The spectral shifts due to charge, i.e. hole accumulation is different from spectral shifts due to surface photo voltage \cite{Haase2000}. The surface photovoltage (SPV) effect is explained in the band bending picture \cite{Demuth1986, Alonso1990}, depending on the doping of semiconductor, i.e. n-type and p-type, where the conduction and valance band edges bend upward and downward the surface, respectively, to equilibrate the Fermi level. The SPV then shifts the electron spectra toward higher binding energies in n-type and lower binding energies in p-type semiconductors.

\begin{figure}
    \centering
    \includegraphics[width=1\linewidth]{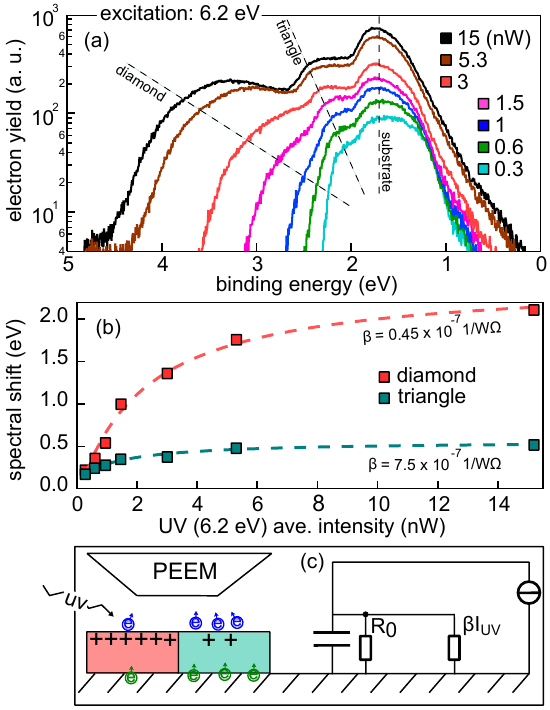}
    \caption{Different crystal domains are charge differently and the more excitation intensity is, the more charge is stored. (a) Overall electron spectra of the crystal for different illumination powers. The dashed lines indicate the sector and substrate specific spectral bands. (b) Spectral shift versus the UV illumination average power, resolved for different sectors. (c) Schematic of single colour experiment. left panel: Red and green rectangles represent the diamond- and triangular-shaped sectors. The blue and green circles represent the photoelectrons and electron supplement from the substrate. The positive charge accumulation near surface is indicated with + sign. right panel: Schematic of RC model with $\text{R}_0$ and $\beta \text{I}_{UV}$ representing the dark resistance of rubrene crystal and resistance induced by UV illumination, respectively.}
    \label{fig:fs_charge}
\end{figure}

\subsection{RC charging model}

As we observed, charging does not occur immediately; instead, it takes time for the crystal sectors to fully charge. Figure \ref{fig:fs_spec_shift}a shows a series of electron spectra taken at different times after illumination, each for 20 seconds. Initially, both sectors share the same spectral position. However, after 60 seconds, the spectrum of the diamond-shaped sectors shifts by nearly 0.6 eV toward higher binding energies. Over time, the spectral shift continues, while the spectrum of the triangular-shaped sectors seems to remain unchanged. This process persists until the spectral shift reaches a saturation point.

We evaluated the charge dynamics for different UV illumination intensity and observed that higher UV power results in stronger spectral shift and charging. Interestingly, the time constant for the charge to reach the saturation, i.e. "fully charged" is independent of the illumination power. Figure \ref{fig:fs_spec_shift}.b is the shift of spectra versus UV illumination time for two different illumination intensities. Both datasets follow the rate equation:

\begin{equation}
     V(t) = V_0 + V_1 \cdot \left(1 - e^{-t/\tau}\right)
\end{equation}

yielding the same time constant of $\approx 2.15$ minutes. Here, V is the spectral shift in eV and $V_0$ and $V_1$ are the offset and maximum spectral shift, respectively. The measured time constant varies across different samples. The charge dynamics follow a rate equation, suggesting the presence of an internal or light-induced resistance that opposes hole accumulation, although the charging process due to hole accumulation remains dominant.

Figure \ref{fig:fs_charge}a shows the overall electron spectra of the crystal, including the diamond- and triangular-shaped sectors. More illumination power leads to more charge of the diamond sector. All of the spectra were taken in a 5-minute data acquisition, well longer than the charging time constant. A slight spectral shift of the triangle sectors is also observed. Figure \ref{fig:fs_charge}b shows the spectral shift of the crystal versus the illumination intensity, resolved for different sectors. 

We describe this spectral shift and charge using the RC model, where hole accumulation due to the UV illumination is considered as the charge stored in the capacitor with capacitance C and resistance R. In figure \ref{fig:fs_charge}c, the left panel is a schematic of the experiment, where the positively charged holes are accumulated differently in different sectors near the surface of the crystal. The diamond- and triangular-shaped sectors are schematically shown as red and green. Green arrows indicate the electron supplement from the substrate, to be replaced by the emitted electrons, indicated by blue arrows. Figure \ref{fig:fs_charge}c right panel shows the corresponding RC model, where $\frac{1}{\text{R}_0}$ is the intrinsic conductance of the rubrene crystal and $\beta I_{UV}$ represents the induced conductance due to generated excitons by the UV lights. Although the generated excitons could increase the conductance, the excitons being generated by UV light (6.2 eV) are expected to be short lived. We believe that a few top layers of the crystal is the active length of the capacitor.

Given $\text{V}=\alpha \text{I} \cdot \text{R}$, with \text{I} being the UV illumination intensity and $\text{R}=\frac{1}{\frac{1}{R_0} + \beta I}$ the equivalent parallel resistance of the crystal dark resistance and the resistance due to electron-hole generation via UV with rate $\beta$, the spectral shift versus the UV illumination intensity could be written as:

\begin{equation}
    V(I_{UV}) = V_0 + \frac{\alpha R_0I_{UV}}{1 + \beta R_0I_{UV}}
\end{equation}

$\alpha=e\frac{dr}{dI_{UV}}$ is proportional to the photoemission rate upon UV illumination with r being the photoelectron count rate with the unit of A/W. $\beta$ has the unit of $1 / \Omega\text{W}$ and indicates the photoconductivity i.e. how efficiently intensity opens conduction channels. The obtained fitting parameters yields $R_0= 3.6 \times10^{15} \Omega$ and $R_0= 1.15 \times10^{15} \Omega$ for diamond- and triangular-shaped sectors, respectively. This explains our observation, in which charge pattern in the crystal remains intact when the illumination is stopped. Accordingly, the photoconductivity factor is obtained $\beta=0.97\times10^{-7} 1/\Omega W$ and $\beta=5.42\times10^{-7} 1/\Omega W$ for diamond- and triangular-shaped sectors.

\begin{figure}
    \centering
    \includegraphics[width=1\linewidth]{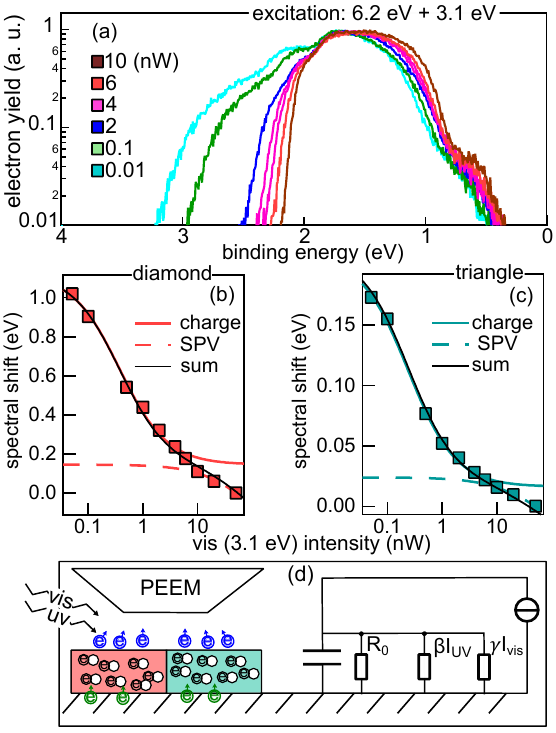}
    \caption{The Vis light can suppress the charge induced by the UV light. (a) Electron spectra of rubrene single crystals illuminated simultaneously by UV (1.4 nW) and different Vis powers. (b and c) Spectral shift versus Vis illumination power, resolved for different sectors. (d) Schematics of double-colour experiment. left panel: Red and green rectangles represent the diamond- and triangular-shaped sectors. The blue and green circles represent the photoelectrons and electron supplement from the substrate. The black and white circles indicate the excitons generated via internal photoelectric upon Vis light. right panel: Schematics of corresponding RC model with $R_0$ representing the dark resistance, $\beta I_{UV}$ and $\gamma I_{vis}$ being the photo resistance due to UV and Vis illumination, respectively.}
    \label{fig:fs_cw}
\end{figure}

Once we apply the Vis light (3.1 eV) simultaneously with the UV light (6.2 eV), the charge pattern neutralizes, instantaneously. Interestingly; the Vis light neutralizes the charge without inducing any further photoemission. Figure \ref{fig:fs_cw}a shows the overall electron spectra of the crystal with UV-vis illumination, at fixed UV intensity (1.4 nW) and different Vis illumination intensities. Each spectra is acquired in 5 minutes. As shown in Figure \ref{fig:fs_cw}a increasing the Vis illumination intensity results in the reduction of spectral shift pointing to charge suppression. At $\text{I}=10 nW$, the charge seems to be neutralized, leading the spectra from the diamond- and triangular-shaped regions to converge (see also Figure \ref{fig:fs_cw_img}b).

Figure \ref{fig:fs_cw}b and \ref{fig:fs_cw}c show the spectral shift versus the Vis light intensity for the diamond- and triangular-shaped sectors, respectively. As shown, the spectra from both sectors tend to shift further toward lower binding energies, even after the charge is fully suppressed. Here we model the situation using two separated approaches. First, for the intensities below $\sim$ 10 nW, the spectral shift is due to charge suppression as a result of high efficient electron-hole generation upon Vis illumination. This process adds another resistance against the hole accumulation. In the contrary of the UV illumination, in the Vis illumination, the exciton generation is rather efficient \cite{Irkhin2012, Naeimi2026} and the singlet fission is the dominant process leading to long lived triplet excitons. The schematics of the RC model is shown in Figure \ref{fig:fs_cw}d right panel. The left panel in Figure \ref{fig:fs_cw}d schematically represents the charge suppression upon Vis illumination. The diamond- and triangular-shaped sectors are represented by red and green and the generated electron holes are represented by black and white circles. Hence, the spectral shift could be expressed as:

\begin{equation}
    V(I_{vis}) = V_0 + \frac{\alpha R_0I_{UV}}{1 + \beta R_0 I_{UV}}(1+\frac{\gamma R_0}{1+\beta R_0 I_{UV}}I_{vis})^{-1}
\end{equation}

With $\gamma$ being the photoconductivity induced by the Vis light. Fitting this equation to the data and using the obtained parameters from the single-colour experiment, yields $\gamma = 0.97\times10^{-6} 1/\Omega \text{W}$ for diamond-shaped and $\gamma = 4.7\times10^{-6} 1/\Omega \text{W}$ for triangle-shaped sectors. The solid coloured curves in Figures \ref{fig:fs_cw}b and \ref{fig:fs_cw}c show the fit. Comparing $\gamma$ and $\beta$, which are photoconductivity factors upon Vis and UV illumination, respectively, the obtained values are in line with the fact that the Vis light generates electron-holes more deficiently than the UV light at least one order of magnitude.

\subsection{Drift diffusion correction}

Once the charge in both sectors is neutralized, the spectral shift due to additional increase in the Vis light intensity, i.e. intensities of the Vis light above $\approx$ 10 nW, could be described by the surface photovoltage, determined by the drift diffusion model:

\begin{equation}
     V(I_{vis}) = V_0 + n \frac{k_B T}{e} \ln(\frac{I_{vis}}{I_0} + 1),
    \label{equ:photovotage}
\end{equation}

where $\frac{\text{K}_BT}{e}=0.025 V$ is the thermal voltage and n is defined as an effective ideality factor that indicates the logarithmic scaling of the photo-induced spectral shift with Vis intensity. The dashed coloured curves in Figures \ref{fig:fs_cw}b and \ref{fig:fs_cw}c, are the fit determined by the drift diffusion model. The extracted ideality factors are $\left| \textbf{n} \right|$=2.75 for diamond- and $\left| \textbf{n} \right|$=0.61 for triangular-shaped sectors. This significant difference of n between the two sectors types, may indicate distinct underlying mechanisms. Ideality factors are defined in a Schottky contact. Having in mind that a very thin conductive oxide layer forms on rubrene crystals \cite{Podzorov2004}, which can take the role as a “metallic” electrode, the ideality factors of the diamond sectors, exceeding the “ideal” situation -that charge carriers stay separated (n$\approx$1)- may be framed as follows: any light or electric field dependent photocurrents or trapping usually tend to increase ideality factors \cite{Wetzelaer2014, Kohler2015}. Trap-assisted (Shockley Read Hall) type of recombination (n=2), we regard as unlikely because the photoemission is not stronger at borders or at domain boundaries. In our luminescence studies \cite{Naeimi2025} we observed distinctly differing photon yield in the sectors, possibly involving loss of charge carriers due to recombination. This carrier leak could drive the ideality factors to values beyond 1.

The solid black curves in Figures \ref{fig:fs_cw}b and \ref{fig:fs_cw}c are the plots representing the sum of both equations with the fit parameters determined independently.

We attribute this light-induced potential change to the formation of a photovoltage, as reported for rubrene \cite{Karak2013, Pandey2007}. The observation of such a photovoltage implies the generation of mobile charge carriers. The typical depletion of charge carriers at a semiconductor surface is associated with an electric field and band bending, contributing to the surface potential. This field extends into the crystal according to the diffusion length of the relevant carriers. Upon illumination, the band bending is reduced, leading to a surface photovoltage that partially restores bulk-like conditions. As noted above, rubrene molecular crystals are highly anisotropic, and so are their surface potentials and corresponding surface photovoltage contributions.

\subsection{Surface potential investigations}

\begin{figure}
    \centering
    \includegraphics[width=1\linewidth]{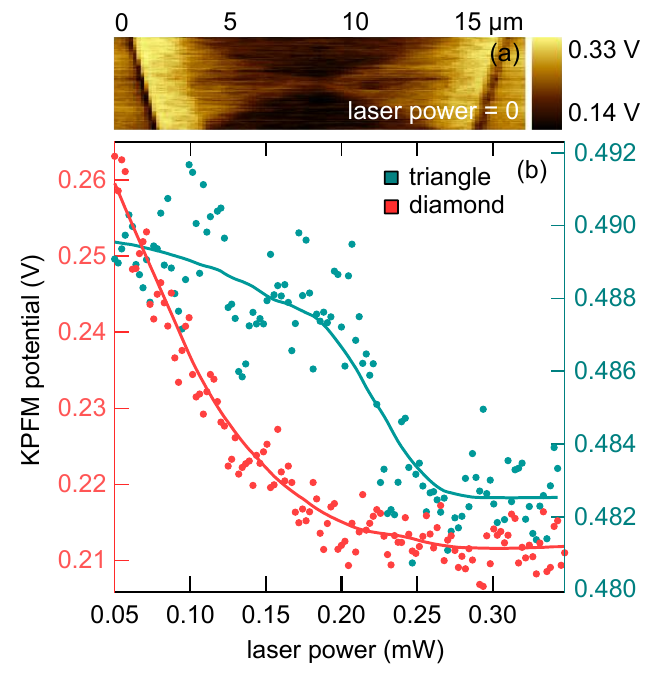}
    \caption{(a) KPFM map of a rubrene crystal showing the contrast of diamond- and triangular-shaped zones, indicating different intrinsic work function of different zones. (b) KPFM potential of a rubrene microcrystal versus grazing illumination intensity, resolved for different sectors.}
    \label{fig:kpfm}
\end{figure}

Regarding that the Surface potential of rubrene orthorhombic crystal is independent of the thickness, we repeatedly measured $\approx$ +300 mV for \textbf{c}-oriented (\textbf{c}-axis perpendicular to the crystal surface) and $\approx$ -150 mV for \textbf{b}-oriented (\textbf{b}-axis perpendicular to the surface) type, without illumination. Surface potential difference due to unit cell orientation is compatible with the common case that more densely packed surfaces exhibit a higher work function: both surface unit cells host two molecules, however the unit cell area of the \textbf{c}-oriented phase is half the area of the \textbf{b}-oriented, i.e., the \textbf{c}-oriented has twice the density of species. However, such density dependent surface potential differences are of the order of 50 ~meV for molecules~\cite{coenen2013}. In our case, the surface potential differences are relatively larger ($\approx 0.5 V$). As shown in Figure \ref{fig:kpfm}a, the crystal sectors (i.e. diamond- and triangular-shaped) also exhibit contrast in the KPFM map, showing potential differences of about 100 mV, as variations in species density arise from the unit cell rotation within each domain. Having in mind, that tensile and compressive strain modulate the pinned Fermi level of rubrene by several 100 meV \cite{Wu2016}, together with the huge anisotropy in expansion coefficient along \textbf{a}- and \textbf{b}- lattice direction (of factor 5), and estimating the temperature during growth at room temperature, such large surface potential differences between crystal types (i.e. \textbf{c}- and \textbf{b}-oriented) and sectors (i.e. diamond- and triangular-shaped) may be attributed to residual stress upon cooling down the film after growth.

Figure \ref{fig:kpfm}b shows the KPFM potential of the diamond- and triangular-shaped sectors of a \textbf{c}-oriented rubrene microcrystal as a function of illumination intensity from a $\sim$532 nm laser applied at a grazing incidence of about 10$^\circ$. A drop in KPFM potential difference—corresponding to an increase in surface potential and charge carrier density (to diminish band bending)—is observed in both sectors. However, the reduction in KPFM potential relative to the laser intensity is significantly more pronounced in the diamond-shaped sectors compared to the triangular-shaped. The respective substrate remains unaffected. The different response to illumination in diamond- and triangular-shaped sectors could be explained by different pinning of the Fermi level at the surface arising from the different defect concentration at the surface due to different unit cell rotation (see Figure \ref{fig:fs_cw_img}).

In the overall picture we explain the charging upon illumination as follows: Under UV excitation (6.2 eV), the one-photon photoemission (1PPE) process liberates electrons similarly efficiently in the diamond-shaped sectors and in the triangular ones. The photoemission leads to enhanced hole residuals near the surface, if the electron supplement from the substrate cannot fill up the holes as effective as the photoemission current. On the other hand, the most effective electron-hole mobility direction is along the \textbf{a}-axis for rubrene orthorhombic crystals\cite{Naeimi2025}. Since the \textbf{a}-axis has more out of plane component in triangular-shaped sectors than in the diamond-shaped, this leads to a different out-of-plane conductivity across the sectors, resulting in greater hole accumulation in the diamond sectors compared to the triangular ones.  This accumulation increases over time characterized by a time constant.

When the Vis light source (3.1~eV) is applied, the generation of triplet excitons via singlet fission becomes the dominant process \cite{Naeimi2026,DiazAndres2024, Ryasnyanskiy2011}. The long-lived triplet excitons create a reservoir of holes, which can be filled by electrons from the substrate and subsequently transferred to the surface to neutralize the accumulated charge. This process occurs rapidly, effectively cancelling the charge pattern. The substantially stronger charging of the diamond sector type implies a lateral electric field across the domain boundaries in the regime of $10^6 V/m$ and eventually an implicit or built-in p-n type junction.

\section{Conclusion}

In this work, we used photoemission electron microscopy and spectroscopy to investigate the charge accumulation in zone-sectored rubrene microcrystals under the influence of the external and internal photoemission process. The resulting charge patterns resemble the diamond- and triangular-shaped sectors, revealing distinct charging behaviours in different zones. Notably, charge accumulation can be effectively suppressed by introducing an additional internal photoelectron excitation using illumination energies below the sample's work function, which is attributed to the generation of electron–hole pairs. The charge dynamics and energetics are described by a RC model and drift diffusion. The compatibility of the drift-diffusion model  with the observed energy shift in the high power regime implies an out of plane surface Schottky contact, as expected for a semiconducting sample. Our findings show that the zone-sectored tabular rubrene single crystals consist of zones with different out-of-plane conductivities and open a route to control and design spatially resolved charge landscapes in organic semiconductor systems. 

\section{Acknowledgement}

Funding by the Deutsche Forschungsgemeinschaft (DFG, German Research Foundation) within projects SFB 1477 “Light-Matter Interactions at Interfaces” (441234705), SFB 1270/2 “Electrically Active Implants” (299150580) and “Application of Interoperable Metadata Standards (AIMS) 2” (432233186) is acknowledged.

\bibliography{apssamp}

@article{Kim2007,
  doi = {10.1016/j.synthmet.2007.05.013},
  url = {https://doi.org/10.1016/j.synthmet.2007.05.013},
  year = {2007},
  month = jun,
  publisher = {Elsevier {BV}},
  volume = {157},
  number = {10-12},
  pages = {481--484},
  author = {Kihyun Kim and Min Ki Kim and Han Saem Kang and Mi Yeon Cho and Jinsoo Joo and Ju Hee Kim and Kyung Hwan Kim and Chang Seop Hong and Dong Hoon Choi},
  title = {New growth method of rubrene single crystal for organic field-effect transistor},
  journal = {Synthetic Metals}
}

@article{Wang2023,
  doi = {10.1002/adfm.202213768},
  url = {https://doi.org/10.1002/adfm.202213768},
  year = {2023},
  month = jan,
  publisher = {Wiley},
  volume = {33},
  number = {14},
  author = {Shu-Jen Wang and Anton Kirch and Michael Sawatzki and Tim Achenbach and Hans Kleemann and Sebastian Reineke and Karl Leo},
  title = {Highly Crystalline Rubrene Light-Emitting Diodes with Epitaxial Growth},
  journal = {Advanced Functional Materials}
}

@article{Zeng2007,
  doi = {10.1016/j.apsusc.2007.01.008},
  url = {https://doi.org/10.1016/j.apsusc.2007.01.008},
  year = {2007},
  month = may,
  publisher = {Elsevier {BV}},
  volume = {253},
  number = {14},
  pages = {6047--6051},
  author = {Xionghui Zeng and Deqiang Zhang and Lian Duan and Liduo Wang and Guifang Dong and Yong Qiu},
  title = {Morphology and fluorescence spectra of rubrene single crystals grown by physical vapor transport},
  journal = {Applied Surface Science}
}

@article{Zhang2010,
  doi = {10.1016/j.orgel.2010.08.019},
  url = {https://doi.org/10.1016/j.orgel.2010.08.019},
  year = {2010},
  month = dec,
  publisher = {Elsevier {BV}},
  volume = {11},
  number = {12},
  pages = {1928--1934},
  author = {Keke K. Zhang and Kejie Tan and Changji Zou and Magnus Wikberg and Laurie E. McNeil and Subodh G. Mhaisalkar and C. Kloc},
  title = {Control of charge mobility in single-crystal rubrene through surface chemistry},
  journal = {Organic Electronics}
}

@article{Finton2019,
  doi = {10.1063/1.5118942},
  url = {https://doi.org/10.1063/1.5118942},
  year = {2019},
  month = sep,
  publisher = {{AIP} Publishing},
  volume = {9},
  number = {9},
  author = {Drew M. Finton and Eric A. Wolf and Vincent S. Zoutenbier and Kebra A. Ward and Ivan Biaggio},
  title = {Routes to singlet exciton fission in rubrene crystals and amorphous films},
  journal = {{AIP} Advances}
}

@article{Ma2012,
  doi = {10.1039/c2cp40449d},
  url = {https://doi.org/10.1039/c2cp40449d},
  year = {2012},
  publisher = {Royal Society of Chemistry ({RSC})},
  volume = {14},
  number = {23},
  pages = {8307},
  author = {Lin Ma and Keke Zhang and Christian Kloc and Handong Sun and Maria E. Michel-Beyerle and Gagik G. Gurzadyan},
  title = {Singlet fission in rubrene single crystal: direct observation by femtosecond pump{\textendash}probe spectroscopy},
  journal = {Physical Chemistry Chemical Physics}
}

@article{AkinKara2022,
  doi = {10.1039/d2cp00985d},
  url = {https://doi.org/10.1039/d2cp00985d},
  year = {2022},
  publisher = {Royal Society of Chemistry ({RSC})},
  volume = {24},
  number = {18},
  pages = {10869--10876},
  author = {Duygu Akin Kara and Edmund K. Burnett and Koray Kara and Ozlem Usluer and Benjamin P. Cherniawski and Edward J. Barron and Burak Gultekin and Mahmut Kus and Alejandro L. Briseno},
  title = {Rubrene single crystal solar cells and the effect of crystallinity on interfacial recombination},
  journal = {Physical Chemistry Chemical Physics}
}

@article{Ye2018,
  doi = {10.1021/acs.chemmater.7b04170},
  url = {https://doi.org/10.1021/acs.chemmater.7b04170},
  year = {2018},
  month = jan,
  publisher = {American Chemical Society ({ACS})},
  volume = {30},
  number = {2},
  pages = {412--420},
  author = {Xin Ye and Yang Liu and Quanxiang Han and Chao Ge and Shuangyue Cui and Leilei Zhang and Xiaoxin Zheng and Guangfeng Liu and Jie Liu and Duo Liu and Xutang Tao},
  title = {Microspacing In-Air Sublimation Growth of Organic Crystals},
  journal = {Chemistry of Materials}
}

@article{Fielitz2016,
  doi = {10.1021/acs.cgd.6b00783},
  url = {https://doi.org/10.1021/acs.cgd.6b00783},
  year = {2016},
  month = jul,
  publisher = {American Chemical Society ({ACS})},
  volume = {16},
  number = {8},
  pages = {4720--4726},
  author = {Thomas R. Fielitz and Russell J. Holmes},
  title = {Crystal Morphology and Growth in Annealed Rubrene Thin Films},
  journal = {Crystal Growth {\&} Design}
}

@article{Euvrard2022,
  doi = {10.1002/adfm.202206438},
  url = {https://doi.org/10.1002/adfm.202206438},
  year = {2022},
  month = oct,
  publisher = {Wiley},
  volume = {32},
  number = {49},
  author = {Julie Euvrard and Oki Gunawan and Antoine Kahn and Barry P. Rand},
  title = {From Amorphous to Polycrystalline Rubrene: Charge Transport in Organic Semiconductors Paralleled with Silicon},
  journal = {Advanced Functional Materials}
}

@article{Jo2015,
  doi = {10.1021/acs.chemmater.5b00884},
  url = {https://doi.org/10.1021/acs.chemmater.5b00884},
  year = {2015},
  month = may,
  publisher = {American Chemical Society ({ACS})},
  volume = {27},
  number = {11},
  pages = {3979--3987},
  author = {Pil Sung Jo and Duc T. Duong and Joonsuk Park and Robert Sinclair and Alberto Salleo},
  title = {Control of Rubrene Polymorphs via Polymer Binders: Applications in Organic Field-Effect Transistors},
  journal = {Chemistry of Materials}
}

@article{Liu2023,
  doi = {10.1002/agt2.347},
  url = {https://doi.org/10.1002/agt2.347},
  year = {2023},
  month = apr,
  publisher = {Wiley},
  volume = {4},
  number = {5},
  author = {Yanping Liu and Xuexiao Yang and Lei Ye and Haibo Ma and Haiming Zhu},
  title = {Molecular stacking controlling coherent and incoherent singlet fission in polymorph rubrene single crystals},
  journal = {Aggregate}
}

@article{Irkhin2012,
  doi = {10.1103/physrevb.86.085143},
  url = {https://doi.org/10.1103/physrevb.86.085143},
  year = {2012},
  month = aug,
  publisher = {American Physical Society ({APS})},
  volume = {86},
  number = {8},
  author = {Pavel Irkhin and Aleksandr Ryasnyanskiy and Marlus Koehler and Ivan Biaggio},
  title = {Absorption and photoluminescence spectroscopy of rubrene single crystals},
  journal = {Physical Review B}
}

@phdthesis{coenen2013,
    author = {Michiel Coenen},
    title = {Combined scanning probe microscopy studies on self-assembled porphyrin monolayers},
    school = {University of Rostock, Rostock, Germany},
    year = {2013}
}

@article{Karak2013,
  title = {Photovoltaic Effect at the Schottky Interface with Organic Single Crystal Rubrene},
  volume = {24},
  ISSN = {1616-3028},
  url = {http://dx.doi.org/10.1002/adfm.201301891},
  DOI = {10.1002/adfm.201301891},
  number = {8},
  journal = {Advanced Functional Materials},
  publisher = {Wiley},
  author = {Karak,  Supravat and Lim,  Jung Ah and Ferdous,  Sunzida and Duzhko,  Volodimyr V. and Briseno,  Alejandro L.},
  year = {2013},
  month = sep,
  pages = {1039–1046}
}

@article{Pandey2007,
  title = {Rubrene/Fullerene Heterostructures with a Half‐Gap Electroluminescence Threshold and Large Photovoltage},
  volume = {19},
  ISSN = {1521-4095},
  url = {http://dx.doi.org/10.1002/adma.200701052},
  DOI = {10.1002/adma.200701052},
  number = {21},
  journal = {Advanced Materials},
  publisher = {Wiley},
  author = {Pandey,  A. K. and Nunzi,  J.‐M.},
  year = {2007},
  month = nov,
  pages = {3613–3617}
}

@Article{Necas2012,
author = {Nečas, David and Klapetek, Petr},
affiliation = {CEITEC — Central European Institute of Technology, Masaryk University Kamenice 753/5, 625 00 Brno, Czech Republic},
title = {Gwyddion: an open-source software for {SPM} data analysis},
journal = {Central European Journal of Physics},
publisher = {Versita, co-published with Springer-Verlag GmbH},
issn = {1895-1082},
pages = {181-188},
volume = {10},
issue = {1},
year = {2012},
doi = {10.2478/s11534-011-0096-2},
}

@article{Ryasnyanskiy2011,
  title = {Triplet exciton dynamics in rubrene single crystals},
  volume = {84},
  ISSN = {1550-235X},
  url = {http://dx.doi.org/10.1103/PhysRevB.84.193203},
  DOI = {10.1103/physrevb.84.193203},
  number = {19},
  journal = {Physical Review B},
  publisher = {American Physical Society (APS)},
  author = {Ryasnyanskiy,  Aleksandr and Biaggio,  Ivan},
  year = {2011},
  month = nov 
}

@article{Bardeen2012,
  title = {Quantum Beats in Crystalline Tetracene Delayed Fluorescence Due to Triplet Pair Coherences Produced by Direct Singlet Fission},
  volume = {134},
  ISSN = {1520-5126},
  url = {http://dx.doi.org/10.1021/ja301683w},
  DOI = {10.1021/ja301683w},
  number = {20},
  journal = {Journal of the American Chemical Society},
  publisher = {American Chemical Society (ACS)},
  author = {Burdett,  Jonathan J. and Bardeen,  Christopher J.},
  year = {2012},
  month = may,
  pages = {8597–8607}
}

@article{Takahashi2019,
  title = {Singlet fission of amorphous rubrene modulated by polariton formation},
  volume = {151},
  ISSN = {1089-7690},
  url = {http://dx.doi.org/10.1063/1.5108698},
  DOI = {10.1063/1.5108698},
  number = {7},
  journal = {The Journal of Chemical Physics},
  publisher = {AIP Publishing},
  author = {Takahashi,  Shota and Watanabe,  Kazuya and Matsumoto,  Yoshiyasu},
  year = {2019},
  month = aug 
}

@article{Wakikawa2025,
  title = {Triplet pair dynamics of singlet fission in orthorhombic polycrystalline powder of rubrene as revealed by magnetoluminescence},
  volume = {162},
  ISSN = {1089-7690},
  url = {http://dx.doi.org/10.1063/5.0251084},
  DOI = {10.1063/5.0251084},
  number = {12},
  journal = {The Journal of Chemical Physics},
  publisher = {AIP Publishing},
  author = {Wakikawa,  Yusuke and Ikoma,  Tadaaki},
  year = {2025},
  month = mar 
}

@article{Clapham2021,
  title = {Beyond single crystals: Imaging rubrene polymorphism across crystalline batches through lattice phonon Raman microscopy},
  volume = {155},
  ISSN = {1089-7690},
  url = {http://dx.doi.org/10.1063/5.0065496},
  DOI = {10.1063/5.0065496},
  number = {23},
  journal = {The Journal of Chemical Physics},
  publisher = {AIP Publishing},
  author = {Clapham,  Margaret L. and Leighton,  Ryan E. and Douglas,  Christopher J. and Frontiera,  Renee R.},
  year = {2021},
  month = dec 
}

@article{Ma2013,
  title = {Fluorescence from rubrene single crystals: Interplay of singlet fission and energy trapping},
  volume = {87},
  ISSN = {1550-235X},
  url = {http://dx.doi.org/10.1103/PhysRevB.87.201203},
  DOI = {10.1103/physrevb.87.201203},
  number = {20},
  journal = {Physical Review B},
  publisher = {American Physical Society (APS)},
  author = {Ma,  Lin and Zhang,  Keke and Kloc,  Christian and Sun,  Handong and Soci,  Cesare and Michel-Beyerle,  Maria E. and Gurzadyan,  Gagik G.},
  year = {2013},
  month = may 
}

@article{Demuth1986,
  title = {Photoemission-Based Photovoltage Probe of Semiconductor Surface and Interface Electronic Structure},
  volume = {56},
  ISSN = {0031-9007},
  url = {http://dx.doi.org/10.1103/PhysRevLett.56.1408},
  DOI = {10.1103/physrevlett.56.1408},
  number = {13},
  journal = {Physical Review Letters},
  publisher = {American Physical Society (APS)},
  author = {Demuth,  J. E. and Thompson,  W. J. and DiNardo,  N. J. and Imbihl,  R.},
  year = {1986},
  month = mar,
  pages = {1408–1411}
}

@article{DiazAndres2024,
  title = {Electronic Couplings for Triplet–Triplet Annihilation Upconversion in Crystal Rubrene},
  volume = {20},
  ISSN = {1549-9626},
  url = {http://dx.doi.org/10.1021/acs.jctc.4c00185},
  DOI = {10.1021/acs.jctc.4c00185},
  number = {10},
  journal = {Journal of Chemical Theory and Computation},
  publisher = {American Chemical Society (ACS)},
  author = {Diaz-Andres,  Aitor and Tonnelé,  Claire and Casanova,  David},
  year = {2024},
  month = may,
  pages = {4288–4297}
}

@article{Naeimi2026,
  title = {Imaging domain boundaries of rubrene thin crystallites by photoemission electron microscopy},
  volume = {279},
  ISSN = {0304-3991},
  url = {http://dx.doi.org/10.1016/j.ultramic.2025.114239},
  DOI = {10.1016/j.ultramic.2025.114239},
  journal = {Ultramicroscopy},
  publisher = {Elsevier BV},
  author = {Naeimi,  Moha and Engster,  Katharina and Pervez,  Waqas and Barke,  Ingo and Speller,  Sylvia},
  year = {2026},
  month = jan,
  pages = {114239}
}

@article{Naeimi2025,
  title = {Zone‐Sectored Organic Crystals with Spatially Resolved Exciton Dynamics},
  ISSN = {2195-1071},
  url = {http://dx.doi.org/10.1002/adom.202502744},
  DOI = {10.1002/adom.202502744},
  journal = {Advanced Optical Materials},
  publisher = {Wiley},
  author = {Naeimi,  Moha and V\"{o}lzer,  Tim and Lange,  Regina and Oldenburg,  Kevin and Lochbrunner,  Stefan and Barke,  Ingo and Speller,  Sylvia},
  year = {2025},
  month = dec 
}

@article{Vollmer2012,
  title = {Two dimensional band structure mapping of organic single crystals using the new generation electron energy analyzer ARTOF},
  volume = {185},
  ISSN = {0368-2048},
  url = {http://dx.doi.org/10.1016/j.elspec.2012.01.003},
  DOI = {10.1016/j.elspec.2012.01.003},
  number = {3–4},
  journal = {Journal of Electron Spectroscopy and Related Phenomena},
  publisher = {Elsevier BV},
  author = {Vollmer,  A. and Ovsyannikov,  R. and Gorgoi,  M. and Krause,  S. and Oehzelt,  M. and Lindblad,  A. and Mårtensson,  N. and Svensson,  S. and Karlsson,  P. and Lundvuist,  M. and Schmeiler,  T. and Pflaum,  J. and Koch,  N.},
  year = {2012},
  month = apr,
  pages = {55–60}
}

@article{Wu2016,
  title = {Strain effects on the work function of an organic semiconductor},
  volume = {7},
  ISSN = {2041-1723},
  url = {http://dx.doi.org/10.1038/ncomms10270},
  DOI = {10.1038/ncomms10270},
  number = {1},
  journal = {Nature Communications},
  publisher = {Springer Science and Business Media LLC},
  author = {Wu,  Yanfei and Chew,  Annabel R. and Rojas,  Geoffrey A. and Sini,  Gjergji and Haugstad,  Greg and Belianinov,  Alex and Kalinin,  Sergei V. and Li,  Hong and Risko,  Chad and Brédas,  Jean-Luc and Salleo,  Alberto and Frisbie,  C. Daniel},
  year = {2016},
  month = feb 
}

@article{Machida2010,
  title = {Highest-Occupied-Molecular-Orbital Band Dispersion of Rubrene Single Crystals as Observed by Angle-Resolved Ultraviolet Photoelectron Spectroscopy},
  volume = {104},
  ISSN = {1079-7114},
  url = {http://dx.doi.org/10.1103/PhysRevLett.104.156401},
  DOI = {10.1103/physrevlett.104.156401},
  number = {15},
  journal = {Physical Review Letters},
  publisher = {American Physical Society (APS)},
  author = {Machida,  Shin-ichi and Nakayama,  Yasuo and Duhm,  Steffen and Xin,  Qian and Funakoshi,  Akihiro and Ogawa,  Naoki and Kera,  Satoshi and Ueno,  Nobuo and Ishii,  Hisao},
  year = {2010},
  month = apr 
}

@article{Nakayama2008,
  title = {Direct observation of the electronic states of single crystalline rubrene under ambient condition by photoelectron yield spectroscopy},
  volume = {93},
  ISSN = {1077-3118},
  url = {http://dx.doi.org/10.1063/1.2998650},
  DOI = {10.1063/1.2998650},
  number = {17},
  journal = {Applied Physics Letters},
  publisher = {AIP Publishing},
  author = {Nakayama,  Yasuo and Machida,  Shinichi and Minari,  Takeo and Tsukagishi,  Kazuhito and Noguchi,  Yutaka and Ishii,  Hisao},
  year = {2008},
  month = oct 
}

@article{Gilbert2000,
  title = {Charging phenomena in PEEM imaging and spectroscopy},
  volume = {83},
  ISSN = {0304-3991},
  url = {http://dx.doi.org/10.1016/S0304-3991(99)00196-5},
  DOI = {10.1016/s0304-3991(99)00196-5},
  number = {1–2},
  journal = {Ultramicroscopy},
  publisher = {Elsevier BV},
  author = {Gilbert,  B and Andres,  R and Perfetti,  P and Margaritondo,  G and Rempfer,  G and De Stasio,  Gelsomina},
  year = {2000},
  month = may,
  pages = {129–139}
}

@article{Wilson2020,
  title = {On space charge effects in laboratory-based photoemission electron microscopy using compact gas discharge extreme ultraviolet sources},
  volume = {22},
  ISSN = {1367-2630},
  url = {http://dx.doi.org/10.1088/1367-2630/abbc29},
  DOI = {10.1088/1367-2630/abbc29},
  number = {10},
  journal = {New Journal of Physics},
  publisher = {IOP Publishing},
  author = {Wilson,  Daniel and Schmitz,  Christoph and Rudolf,  Denis and Wiemann,  Carsten and Schneider,  Claus M and Juschkin,  Larissa},
  year = {2020},
  month = oct,
  pages = {103019}
}

@article{Haase2000,
  title = {Surface photovoltage imaging for the study of local electronic structure at semiconductor surfaces},
  volume = {19},
  ISSN = {1366-591X},
  url = {http://dx.doi.org/10.1080/01442350050020897},
  DOI = {10.1080/01442350050020897},
  number = {2},
  journal = {International Reviews in Physical Chemistry},
  publisher = {Informa UK Limited},
  author = {Haase,  G.},
  year = {2000},
  month = apr,
  pages = {247–276}
}

@article{Alonso1990,
  title = {Surface photovoltage effects in photoemission from metal-GaP(110) interfaces: Importance for band bending evaluation},
  volume = {64},
  ISSN = {0031-9007},
  url = {http://dx.doi.org/10.1103/PhysRevLett.64.1947},
  DOI = {10.1103/physrevlett.64.1947},
  number = {16},
  journal = {Physical Review Letters},
  publisher = {American Physical Society (APS)},
  author = {Alonso,  M. and Cimino,  R. and Horn,  K.},
  year = {1990},
  month = apr,
  pages = {1947–1950}
}

@article{Li2019,
  title = {Photoemission spectroscopy of rubrene thin films doped with heavy alkali metal: A first-principles investigation},
  volume = {132},
  ISSN = {0022-3697},
  url = {http://dx.doi.org/10.1016/j.jpcs.2019.03.026},
  DOI = {10.1016/j.jpcs.2019.03.026},
  journal = {Journal of Physics and Chemistry of Solids},
  publisher = {Elsevier BV},
  author = {Li,  Tsung-Lung and Lu,  Wen-Cai},
  year = {2019},
  month = sep,
  pages = {1–9}
}

@article{Wetzelaer2014,
  title = {Diffusion-driven currents in organic-semiconductor diodes},
  volume = {6},
  ISSN = {1884-4057},
  url = {http://dx.doi.org/10.1038/am.2014.41},
  DOI = {10.1038/am.2014.41},
  number = {7},
  journal = {NPG Asia Materials},
  publisher = {Springer Science and Business Media LLC},
  author = {Wetzelaer,  Gert-Jan A H and Blom,  Paul W M},
  year = {2014},
  month = jul,
  pages = {e110–e110}
}

@BOOK{Kohler2015,
  title     = "Electronic processes in organic semiconductors",
  author    = "Kohler, Anna and Bassler, Heinz",
  publisher = "Wiley-VCH Verlag",
  month     =  apr,
  year      =  2015,
  address   = "Weinheim, Germany",
}

@article{Podzorov2004,
  title = {Light-induced switching in back-gated organic transistors with built-in conduction channel},
  volume = {85},
  ISSN = {1077-3118},
  url = {http://dx.doi.org/10.1063/1.1836877},
  DOI = {10.1063/1.1836877},
  number = {24},
  journal = {Applied Physics Letters},
  publisher = {AIP Publishing},
  author = {Podzorov,  V. and Pudalov,  V. M. and Gershenson,  M. E.},
  year = {2004},
  month = dec,
  pages = {6039–6041}
}

@article{Yang2018,
  title = {Polymer/Si Heterojunction Hybrid Solar Cells with Rubrene:DMSO Organic Semiconductor Film as an Electron-Selective Contact},
  volume = {122},
  ISSN = {1932-7455},
  url = {http://dx.doi.org/10.1021/acs.jpcc.8b07987},
  DOI = {10.1021/acs.jpcc.8b07987},
  number = {41},
  journal = {The Journal of Physical Chemistry C},
  publisher = {American Chemical Society (ACS)},
  author = {Yang,  Linlin and Chen,  Jianhui and Ge,  Kunpeng and Guo,  Jianxin and Duan,  Qingchun and Li,  Feng and Xu,  Ying and Mai,  Yaohua},
  year = {2018},
  month = sep,
  pages = {23371–23376}
}

\end{document}